# Effect of Pt bottom electrode texture selection on the tetragonality and physical properties of Ba0.8Sr0.2TiO3 thin films produced by pulsed laser deposition

J. P. B. Silva,1 K. C. Sekhar,1 A. Almeida,2 J. Agostinho Moreira,2 J. Martín-Sánchez, 1,3 M. Pereira,1 A. Khodorov,1 and M. J. M. Gomes1

1 Centre of Physics, University of Minho, Campus de Gualtar, 4710-057 Braga, Portugal

2 Departamento de Fısica e Astronomia, IFIMUP and IN-Institute of Nanoscience and Nanotechnology, Faculdade de Ciencias da Universidade do Porto, Rua do Campo Alegre 687, 4169-007 Porto, Portugal

3 Laser Processing Group, Instituto de Optica, CSIC, C/Serrano 121, 28006 Madrid, Spain

**ABSTRACT**

The effect of platinum (Pt) bottom electrode texture on the tetragonality, dielectric, ferroelectric, and polarization switching response of pulsed laser deposited Ba0.8Sr0.2TiO3 (BST) thin films has been studied. The x-ray diffraction and Raman analysis revealed the higher tetragonality of BST films when they were grown on higher (111) textured Pt layer. The properties like dielectric permittivity, polarization, switching time, and leakage currents were found to be correlated to tetragonality and orientation of the BST films. The polarization current was observed to be higher in BST films on Pt epitaxial layer and it exhibits exponential dependence on the electric field. The voltage-current measurements displayed Ohmic behavior of leakage current irrespective of Pt texture for low voltages (up to 1 V), whereas at higher voltages the conduction mechanism was found to be dependent on texture selection of bottom Pt electrode.

## 1. Introduction

Ferroelectric thin films are promising candidates for various applications such as ferroelectric memories, infrared sensors, tunable microwave devices, and other integrated technologies.1

Among many ferroelectrics, barium strontium titanate (Ba$_x$Sr$_{1x}$TiO$_3$) is an interesting material for device applications due to their high dielectric constant, relatively low losses, and fast switching speed over a wide frequency range. Moreover, these properties can be tailored for specific applications by controlling the barium to strontium ratio. A wide variety of techniques such as sol-gel, metal-organic chemical vapour deposition, RF magnetron sputtering, and laser ablation have been employed to fabricate Ba$_{0.8}$Sr$_{0.2}$TiO$_3$ (BST) thin films and their electrical properties were extensively investigated.[2–4] In fact, the properties of BST thin films have been found to depend substantially on the fabrication process, microstructure, and film thickness.[3,5,6] Many researchers have continuously made attempts to enhance the properties of BST thin films by improving the crystalline quality, film-electrode interface, and interfacial strain.[7,8] Recently, it has been shown that textured BST thin films can be grown using the single crystal substrates like MgO and SrTiO$_3$[9,10] and very thin buffered layer like SrRuO$_3$ or (La,Sr)MnO$_3$ as a bottom electrode.[7,11] The studies on textured BST thin films revealed that their ferroelectric properties strongly depend upon texture selection.[12] Hence, the control of crystal orientation is the easy way to tailor the properties as desired for devices. However, the control of texture selection of BST thin films on the conventional metal electrodes like Pt has not been much studied yet, whereas it is important from technological point of view. It is also observed that bottom electrode directly influences the properties of BST films. However, it seems that there are no reports showing the influence of the Pt layer texture selection on the properties of the BST films. Therefore, we made an attempt to study the effect of platinum (Pt) bottom electrode texture selection on the tetragonality and physical properties of BST thin films. In view of this, the Ba$_{0.8}$Sr$_{0.2}$TiO$_3$ thin films were grown using the pulsed laser deposition technique on epitaxial, highly and less (111) textured Pt layers integrated with Si substrate and their physical properties were investigated. The properties of Ba$_{0.8}$Sr$_{0.2}$TiO$_3$ thin films, such as dielectric constant, leakage current, remnant polarization, and switching time, were found to be closely related to the tetragonality caused by Pt bottom electrode layer texture selection.

**2. Experiment**

The target with composition of BST was prepared by conventional solid state reaction route using raw materials of BaCO3 (Aldrich, 99%), SrCO3 (Aldrich, 98%), and TiO2 (Merck, > 99%), and then, sintered at 1250 C. The films were grown by pulsed laser deposition using the 248 nm line of an excimer with a frequency of 10 Hz and energy density of 5 J/cm2 was focused on the target. The deposition was performed at 700 C in an oxygen gas pressure of 0.1 mbar on commercially available platinized Si substrates with different Pt texture selection. The substrates A and C are of Neyco and B is of Radiant manufactures. The substrates A and B are SiO2/TiO2/Pt layers and substrate B is SiO2/Ti/Pt layers grown on (100) Si. Subsequently, an annealing was performed in air at 700 C for 30 min in order to improve the crystalline structure. The thickness of films was found to be ~400 nm. The BST films and substrates were structurally characterized using a Philips x-ray diffractometer (model PW1710), and the CuKa radiation ($\lambda$=0.154 nm). The surface morphology of BST films and platinized silicon substrates was scanned with a commercial Nanoscope III AFM system in tapping mode. The unpolarized Raman spectra of BST films have been recorded in the backscattering geometry, at room temperature, by using an Olympus microscope and a 100x objective. The 514.5 nm polarized line of an Arþ laser was used for excitation, with an incident power of about 16 mW impinging on the sample. The scattered light was analyzed using a T64000 Jobin-Yvon spectrometer, operating in triple subtractive mode, and equipped with liquid nitrogen cooled CCD detector in a Stokes frequency range from 200 to 800 cm1 . Gold (Au) electrodes, with diameter of 1 mm, were deposited by thermal evaporation on the top of BST films and then annealed at 200 °C for 30 min in air to improve the electrode-film interface. The frequency dependence dielectric permittivity measurements were performed with QuadTech 7600b Precision LCR Meter at high measurements accuracy (±0.05%), and at an applied voltage of 30 mV. The ferroelectric hysteresis loops (P-E) were traced by a modified Sawyer-Tower circuit using a sinusoidal signal at a frequency of 1 kHz. The polarization reversal characteristics have been studied by applying square pulse, and the corresponding current was measured across a resistance of 100 X connected in series with the samples. The leakage current of Pt/BST/Au capacitors was measured using a Keithley 617 programmable electrometer.

## 3. Results and discussion

**A. Structural and morphological characterization of Pt substrates.**

Figure 1 shows the x-ray diffraction (XRD) patterns of different commercial platinized substrates, used for the film deposition. As depicted in Fig. 1(a), XRD pattern of first type of Pt/TiO2/SiO2/Si substrate shows no peaks of Pt layer but (111) and this is in an agreement with manufacturer that this substrate (for convenience substrate A or epitaxial substrate) has the epitaxial Pt layer. In the studied range, the XRD pattern of other Pt/Ti/SiO2/Si and Pt/TiO2/SiO2/Si substrates consists of two peaks such as (111) and (200) related to Pt layer. The ratio of relative intensities of these two peaks was found to be 0.07 and 0.20 and the full width of half maximum (FWHM) was equal to 0.20 and 0.42 for Pt/TiO2/ SiO2/Si (substrate B) and Pt/Ti/SiO2/Si (substrate C) substrates, respectively. The ratio I(200)/I(111) is ——— 0.5 for polycrystalline Pt (JPCDS card no. 4-0802), hereafter, the respective substrates are highly (111) (substrate B) and less (111) textured (substrate C, with poor crystalline properties of Pt layer). Figures 2(a)–2(c) show the atomic force microscopic images of the A, B, and C substrates. The substrates A and C consist of homogenous Pt grains with mean size of 30 nm. However, the grain structure was found to be denser in the case of epitaxial substrate (Fig. 2(c)) than for substrate C. The substrate B has non-homogenous distribution of grains and also has some clusters of Pt grains with size of about 180 nm. The average root mean square (RMS) value of the surface roughness lies between 2 and 4 nm in all the substrates which ensures a quite planar interface between the Pt layer and the BST film.

**B. Structural properties of BST thin films.**

**1. X-ray diffraction and AFM studies.**

The XRD patterns of BST thin films deposited on A, B, and C substrates are illustrated in Fig. 3(a). All the BST thin films have been crystallized. The presence of clear splitting of (200) peak around $2\theta=45°$ in the film grown on epitaxial substrate A confirms its tetragonal symmetry at room temperature. The extended scan of XRD around $2\theta=45°$ is shown in Fig. 3(b). The splitting of the (200) peak is not visible in the case of film grown on substrate B, but this peak is broader and asymmetric as compared with the film grown on substrate C. Hence, the decomposition of the peak was performed assuming the existence of two (002) and (200) peaks of tetragonal phase as shown in Fig. 3(b). The observed shift in the diffraction peak positions of the films with respect to bulk BST gives a small change in the lattice parameters. The lattice parameters calculated by the least-square method, tetragonality (c/a ratio), and the cell volume of thin films grown on different substrate are shown in the Table I. The tetragonality ratio is found to be the highest in the films grown on epitaxial Pt layer. The c/a ratio obtained for the films grown on epitaxial substrate is similar to the values found in literature for bulk BST[13–15] and is slightly lower than the values obtained in the BST films grown on MgO substrate.[16] For a qualitative characterization of the films orientation, the following equation was used:[17]

$$\frac{I'}{I'_{100} + I'_{101} + I'_{111}}, \qquad (1)$$

where $I'$ is equal to $I/I^*$. The $I$ and $I^*$ are the intensities of a particular reflection in the film and powdered polycrystalline (JPCDS card No. 44-93), respectively. The intensity ratios are shown in Table I. It is seen that the films become more (100) textured with the increase of degree of

(111) texture of Pt layer. The AFM images of the BST thin films grown on different Pt layers reveal the similar morphology of BST thin films irrespective of Pt orientation as shown in Fig. 4. All the films consist of homogenous distribution of grains with an average size of 30 nm. The average RMS value of surface roughness was found to be 6–7 nm.

| Samples | a(Å) | c(Å) | c/a | Volume(Å$^3$) | $\frac{I'_{100}}{I'_{100}+I'_{101}I'_{111}}$ | $\frac{I'_{101}}{I'_{100}+I'_{101}I'_{111}}$ | $\frac{I'_{111}}{I'_{100}+I'_{101}I'_{111}}$ |
|---|---|---|---|---|---|---|---|
| BST/A substrate | 3.984 | 4.011 | 1.007 | 63.7 | 0.78 | 0.22 | 0.00 |
| BST/B substrate | 3.981 | 4.002 | 1.005 | 63.5 | 0.12 | 0.68 | 0.20 |
| BST/C substrate | 3.991 | 3.999 | 1.002 | 63.7 | 0.09 | 0.65 | 0.26 |

TABLE I. Lattice parameters, tetragonality (c/a), cell volume, and the relative intensity ratio of peaks in BST thin films grown on different substrates.

**2. Raman spectroscopy.**

Figure 5 shows the room temperature unpolarized microRaman spectra of sintered target (inset on Fig. 5) and thin films grown on different substrates in the spectral range of 200–800 cm$^{-1}$. The Raman spectra of target consists of a broad peak centered at 235 cm$^{-1}$, a weak shoulder peak at 305 cm$^{-1}$, an asymmetric peak near 519 cm$^{-1}$, and a broad weak peak at 720 cm$^{-1}$. The peaks at 235, 305, 519, and 720 cm$^{-1}$ can be assigned to A1 (2TO), E (3TO + 2LO) + B1, A1 (3TO), and E (4LO) phonon modes, respectively.[18–20] The presence of peaks at 305 and 720 cm1 in Raman spectra of the target evidences its tetragonal symmetry. The Raman spectra of films grown on different substrates consist of the referred above peaks, but their positions are slightly shifted as compared to the target. In order to analyze them in detail, a sum of independent damped oscillators, according to the general formula[21]

$$I(\omega,T) = (1+n(\omega,T))\sum_{j=1}^{N} A_{oj} \frac{\omega \Omega_{oj}^2 \Gamma_{oj}}{(\Omega_{oj}^2-\omega^2)^2+\omega^2\Gamma_{oj}^2} \quad (2)$$

was fitted to the experimental spectra. In Eq. (2), $n(\omega,t)$, is the Bose-Einstein factor, A$_{oj}$, Ω$_{oj}$, and Γ$_{oj}$ are the strength, wave number, and damping coefficient of the jth oscillator, respectively. The solid line in Fig. 6 shows the resultant fitting curve to the experimental data (discrete points). The wave number of the Raman active modes observed in the films is summarized in Table II. From the fitting, it can be seen three more bands in the 450–700 cm$^{-1}$ spectral range, which are marked with *. These bands are recognised as disorderactivated bands[16,22] arising from the random distribution of Ti ions, which can occupy four off-center sites in the tetragonal phase. The Ba/Sr substitution could cause local distortions, partially breaking the translational symmetry of the BT lattice. Also from the Fig. 6, it is clear that the intensity of these peaks is higher for the film deposited on the less (111) textured substrate. This feature reveals that this Pt layer enhances disorder in the films, resulting in a decrease of tetragonality. The intensities of Raman modes E (3TO + 2LO + B1 and A1 (2TO) and the ratio of their intensities were also shown in Table II. These results also suggest that tetragonality increases for more (111) textured Pt layer and is in a good agreement with XRD analysis.

**C. Electrical properties.**

**1. Dielectric and ferroelectric properties.**

Figure 7 shows the frequency dependent dielectric permittivity (r) and dielectric loss (tan d) of BST films grown on different Pt layers. Both r and tan d have shown a small dispersion with frequency. At high frequencies, the values of r and tan d are found to be nearly the same in the films grown on epitaxial and highly (111) textured Pt layers. Nevertheless, the BST thin films grown on the less (111) textured Pt layer (substrate C) have much low r (nearly 3 times) and high tan d in the all measured frequency range. The r and tan d values obtained in the films are higher as compared to that of values reported by Panda et al.[23] and are similar to other BST films grown on platinized substrates.[12,24,25] Figure 8 shows the P-E hysteresis loop of BST thin films grown on different Pt layers. These loops were obtained with applying sinusoidal signal of frequency 1 kHz at room temperature. The values of remnant polarization (Pr), maximum polarization (Pmax), and coercive field (Ec) observed in different films are shown in

Table III. The films produced on epitaxial Pt layer (substrate A) exhibited hysteresis loops with Pr of 0.5 µC/cm2 , Pmax of 7.8 µC/cm2 at 275 kV/cm, and Ec of 16.0 kV/cm. The Pr value is found to be in a good agreement with the values obtained in BST films produced by other techniques.12,26 Whereas, the polarization values observed for the films grown on substrate C is much lower as compared to the values observed for the films deposited on A and B substrates and this can be attributed to the value of tetragonality. The value of the (c-a)/a ratio (%) in the films grown on A, B, and C Pt layers was found to be 0.7%, 0.5%, and 0.2%, respectively. This is in consistent with the fact observed by Golovko et al.27 that BST ceramics exhibit hysteresis loops only when the ratio (%) of (c-a)/a should be ≥0.19%.

**2. Domain reversal process.**

The time dependence polarization reversal characteristics of the BST thin films grown on different substrates is shown in Fig. 9. The bell shape curve in polarization current arises due to domains reversal. When an electric field is reversed, the domain reversal process takes place via nucleation, growth, and coalescence of domains.28 The whole process causes a rapid change in the polarization current. The polarization current at any time is proportional to the rate of change in fraction of domain reversal volume to the total volume of domains in the thin films. The maximum value in polarization current ($i_m$) occurs at which most of the domains are reversed and is found to be enhanced in the films grown on epitaxial substrates. The polarization during the switching (Ps) can be estimated from the area (equal to charge Q) under the transient curves using the relation Q= 2$P_s$A (A is the area of electrode) was found to be in a good agreement with the values obtained from the P-E loops29 (Table III). The effect of pulse amplitude on the time dependence of polarization current transients of BST films grown on epitaxial substrate is shown in Fig. 10. As the pulse amplitude increases from 3 to 11 V, the im value increases from 24 to 69 mA, whereas the tm values decrease from 0.61 to 0.25 µs. This could be due to switching of new regions into the field direction at higher pulse amplitudes.

The $i_m$ values exhibit exponential dependence on the 1/E according to equation given as follows:[30]

$$i_m = i_0 \exp(-\alpha_i/E), \qquad (3)$$

where $\alpha_i$ is the activation field. The semi-log plot of $i_m$ versus 1/E is shown in the inset of Fig. 10. The activation field estimated from Eq. (3) is found to be 74.7 kV/cm.

**3. Leakage current characteristics.**

Because in many applications the ferroelectrics operate under applied external electrical field, the study of leakage current mechanism is very important. Figure 11 depicts the leakage current versus applied voltage (I–V) characteristics of the BST thin films grown on different substrates. We will define forward bias as the case when the top Au electrode is at a higher potential than the bottom Pt electrode and because we are interested to study the effect of Pt bottom electrode texture, the reverse bias currents will be analyzed. The I–V characteristics were found to be slightly asymmetric, which may originate from the different work functions of Pt and Au electrodes which are about 5.6 and 5.40 eV, respectively, depending on crystallographic direction.[31] In the voltage region (4–5 V), it can be seen that the leakage current increases when the Pt orientation changes from the epitaxial to less textured ones. In order to understand the conduction mechanism in BST thin films, several leakage models, such as the space-charge-limited current (SCLC), Poole-Frenkel (PF), Schottky emission (SE), and the hopping conduction, were considered.[32–34] At low voltages (0–1 V), the leakage current was found to follow the Ohmic behavior, irrespective of substrate used. In the voltage range (1–5 V), the I–V dependence becomes non-linear that will be discussed below.

| Samples | $A_1(2TO)$ (cm$^{-1}$) | $E(3TO+2LO)+B_1$ (cm$^{-1}$) | $A_1(3TO)$ (cm$^{-1}$) | $E(4LO)$ (cm$^{-1}$) | Disorder-induced bands (cm$^{-1}$) | $I_{A_1(2TO)}$ | $I_{E(3TO+2LO)+B_1}$ | $I_{E(3TO+2LO)+B_1}/I_{A_1(2TO)}$ |
|---|---|---|---|---|---|---|---|---|
| Target | 235 | 305 | 519 | 720 | ... | ... | ... | ... |
| BST/A substrate | 245 | 300 | 515 | 730 | 589, 612, 634 | 2946 | 280 | 0.095 |
| BST/B substrate | 245 | 300 | 515 | 731 | 589, 612, 634 | 1707 | 158 | 0.093 |
| BST/C substrate | 240 | 300 | 513 | 730 | 589, 612, 634 | 2791 | 250 | 0.090 |

TABLE II. Optical phonon frequencies (in cm$^{-1}$) of Raman modes observed in BST target and in films grown on different substrates, the intensity of A1 (2TO) and E (3TO + 2LO) + B1 modes and also their intensity ratio in BST films.

In the case of BST film deposited on substrate C, the leakage current could be well scaled with $J \sim E^2$, which relates to the bulk limited SCLC mechanism.34 However, as is seen from Fig. 12, the measurements on films of different thicknesses revealed an increase of current density with the thickness that speaks in favor of the hopping conduction mechanism.35,36 In this case, the carrier injection in the film is interface controlled and the movement of the injected carriers inside the film is through a hopping mechanism in a narrow band located in the gap and associated with some kind of structural defects. According to the hopping conduction theory, the current density is described by35,36

$$J \sim \sinh\left(\frac{qEa}{2kT}\right)\exp\left(\frac{W_a}{kT}\right), \qquad (4)$$

where a is the distance between the nearest neighbors, q the electronic charge, T the temperature, E the electrical field, k the Boltzmann constant, and Wa the activation energy for the hopping mechanism. The plot of J versus sinh(αV) (inset to Fig. 12) shows that Eq. (4) fits well the experimental data, by adjusting the parameter α = (qa)/(2kTw) and the fixed a = 4 A°, where w is the thickness of the layer over which the voltage drop is equal to the applied voltage V. The best fit leads to a thickness w of 8.0 nm. This value is much lower than the film thickness (400 nm) that allows us to suggest a partial depletion of the film. These results are in a good agreement with the study of BaTiO3 films36 where the depletion region was suggested to be about 15 nm at room temperature. In the case of BST film grown on epitaxial substrate A, the data showed a good fitting to two conduction mechanisms, namely PF and Schottky emission. The former is bulk limited that consists of field-assisted hopping from one defect to another. The later is the interface limited and caused by thermionic injection of carriers from metal electrode into ferroelectric. According to PF emission, the current density is as follows:32

$$J = AE \exp\left[\frac{-q(\varphi_t - \sqrt{qV/\pi\varepsilon_0\varepsilon w})}{kT}\right] \quad (5)$$

and according to Schottky emission, the current density is given by33

$$J = A^{**}T^2 \exp\left[\frac{-q(\varphi_B - \sqrt{qV/4\pi\varepsilon_0\varepsilon w})}{kT}\right], \quad (6)$$

where J is the leakage current density, A is a constant, E is the electric field, $u_t$ is the trapped level, A** is the effective Richardson constant, T is the temperature, q is the electronic charge, $u_B$ is the Schottky barrier height, V is the applied voltage, $e_0$ is the free space dielectric constant, e is the optical dielectric constant, w is the depletion layer width, and k is the Boltzmann constant. As it emerges from Eqs. (5) and (6), these conduction mechanisms have similar functional dependence of current on applied voltage and one of the ways to distinguish between them is the estimated value of dielectric permittivity. Since PF conduction is a bulk mechanism, it is usual to assume w to be equivalent to the film thickness ~400 nm. In the case of Schottky model (Eq. (6)), we examined different situation from full depletion to the values of w down to few nanometers.37 However, data handling in both cases of PF and Schottky emission leads to non-physical values of e. The "standard Schottky equation" relates to the situation when the mobility is high and transport is limited by the recombination velocity at the potential barrier maximum. In the case of reduced mobility (that could be in our case of polycrystalline film due to scattering on grain boundaries), more appropriate may be the use of Simmons expression (7)38

$$J(T,E) = \alpha T^{3/2} E\mu \left(\frac{m^*}{m_0}\right)^{3/2} \exp\left(-\frac{\phi_b}{KT}\right) \exp\left(\beta\sqrt{E}\right), \quad (7)$$

which takes into account both the carrier density at the potential barrier maximum near the interface and the mobility which is a bulk property. As is seen from Fig. 13, the data can be well fitted with the Simmons equation and give the appropriate values of ε = 3.90 and uB ¼ 0.50 eV, which are in a good agreement with the reported values for BST materials.39,40 In the case of BST film grown on substrate B, none of the mentioned models fit the experimental data in a wide voltage range. As can be seen in Fig. 14, at high voltages (2.8–5 V), the data fit to the "standard Schottky equation" (6) with e ¼ 3.99 and uB ¼ 0.36 eV and suggesting full depletion. In the middle range (1–2.8 V), the most probably, several mechanisms give a considerable contribution to the leakage current.

| Samples | $P_r$ ($\mu C/cm^2$) | $E_c$ (kV/cm) | $P_{max}$ ($\mu C/cm^2$) from P-E loops | $P_{max}$ ($\mu C/cm^2$) from switching current |
|---|---|---|---|---|
| BST/A substrate | 0.5 | 16.0 | 7.8 | 8.3 |
| BST/B substrate | 0.3 | 16.0 | 6.7 | 6.2 |
| BST/C substrate | 0.03 | 29.0 | 0.3 | 0.4 |

TABLE III. Polarization values and coercive fields observed in the BST thin films grown on different substrates.

So, we can conclude that BST films grown in the same conditions on platinized substrate of different texture showed the different conduction mechanisms. This is due to the texture selection of Pt layer. As it known, the work function of Pt depends on crystallographic direction. This is 5.93 eV for ⟨111⟩ direction and decreases to 5.64 eV in the case of polycrystalline Pt.31 Thereby, when BST films were grown on epitaxial (111) Pt layer (substrate A), the barrier height for the injection of carriers from Pt electrode into BST film was maximum and the conductivity was governed by both interface and bulk properties (Simmons model). In the case of less textured Pt layers (substrates B and C), the barrier height was lower, the carriers were more easily injected into film, and conductivity was governed by bulky properties of BST films.

**IV. CONCLUSIONS.**

This work highlights the strong relationship between the texture of Pt layer and the tetragonality of the grown BST thin films. A detailed study of the influence of the bottom Pt electrode texture on the structural and ferroelectric properties of BST thin films produced by pulsed laser deposition has been performed. The XRD and Raman analysis suggest that films on epitaxial Pt layer have the highest tetragonality, which causes superior dielectric and ferroelectric properties than compared to the films on textured Pt layers. The polarization estimated from P-E loops agrees well with the one calculated from the domain reversal process. The polarization current exhibits exponential dependence on the pulse amplitude. In the case of BST film grown on less textured Pt layer, the leakage current was observed to be governed by the hopping conduction mechanism, whereas in the case of epitaxial Pt bottom electrode, the Schottky-Simmons emission was responsible for the leakage current at high fields..

**Acknowledgements**

The author J.P.B.S. thanks FCT for the financial support (Grant SFRH/BD/44861/2008). K.C.S. thanks FCT for Postdoc Grant (SFRH/BPD/68489/2010). J.M.S. thanks FCT for the financial support (Grant SFRH/BPD/64850/2009).

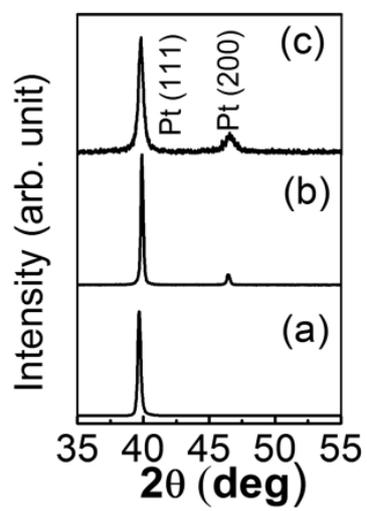

FIG. 1. XRD patterns of substrates (a) A, (b) B, and (c) C.

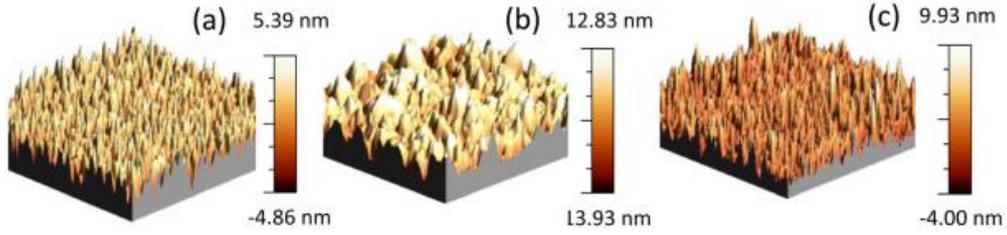

FIG. 2. AFM images of substrates (a) A, (b) B, and (c) C (1 lm 1 lm).

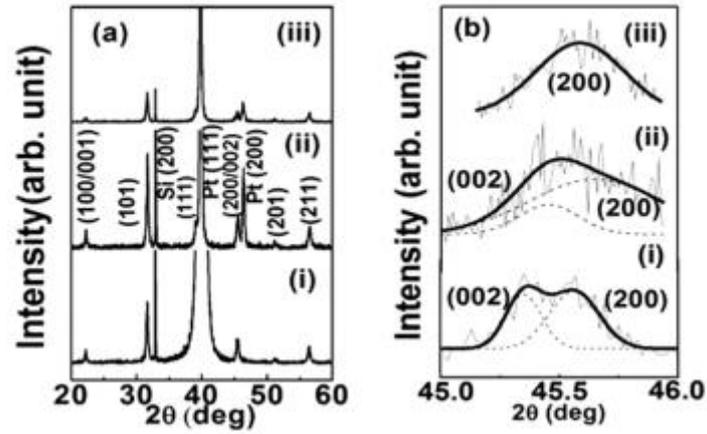

FIG. 3. XRD patterns of BST thin films grown on substrates (i) A, (ii) B, (iii) C and (b): extended XRD scans.

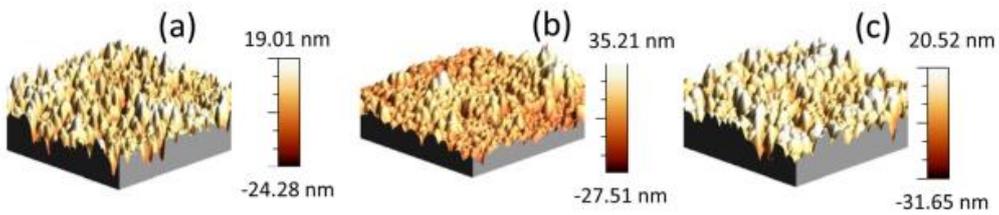

FIG. 4. AFM images of BST thin films grown on substrates (a) A, (b) B, and (c) C (1 μm x 1 μm).

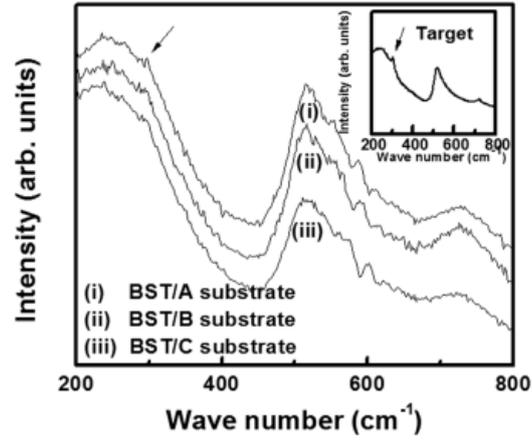

FIG. 5. Raman spectra of BST thin films grown on different substrates and the BST target (inset). The arrows evidence the tetragonal peak.

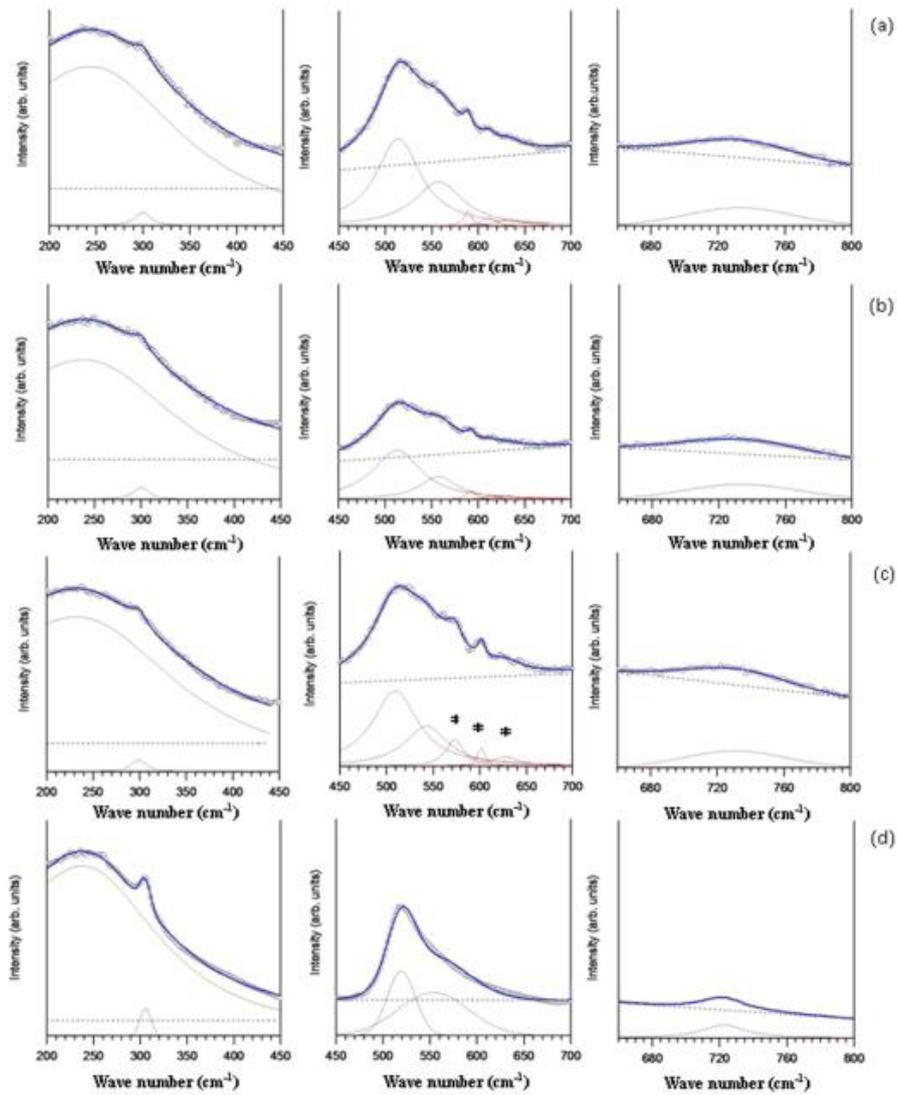

FIG. 6. Fitting of Eq. (2) to the Raman spectra of BST thin film grown on substrates (a) A, (b) B, (c) C, and (d) BST target.

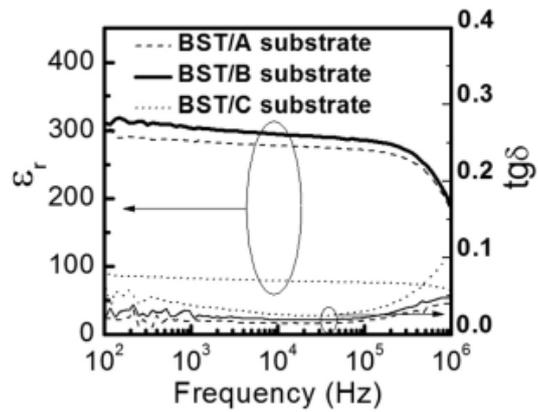

FIG. 7. Frequency dependent dielectric constant and tangent loss of BST films grown on different substrates.

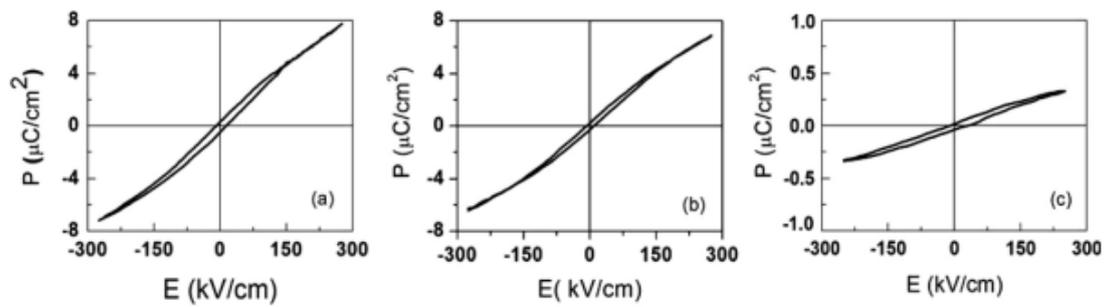

FIG. 8. P-E hysteresis loops of BST films grown on substrates (a) A, (b) B, and (c) C.

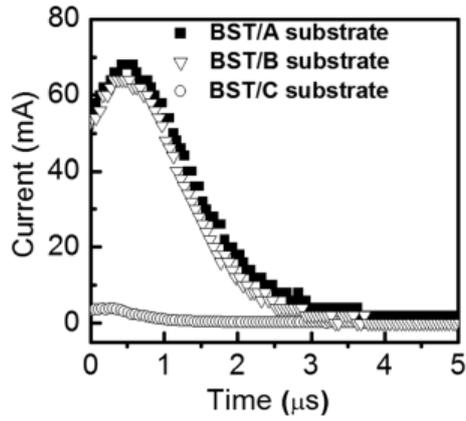

FIG. 9. Time dependence polarization reversal characteristics of BST films on different substrates measure at pulse amplitude of 11 V

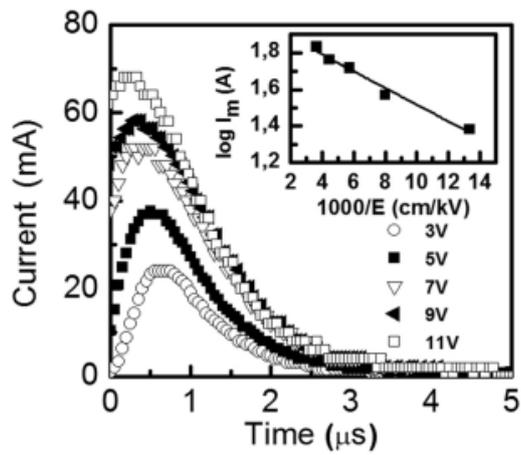

FIG. 10. Effect of pulse amplitude on polarization reversal current of BST films on substrate A. The inset shows semi-log plot of $i_m$ versus 1/E.

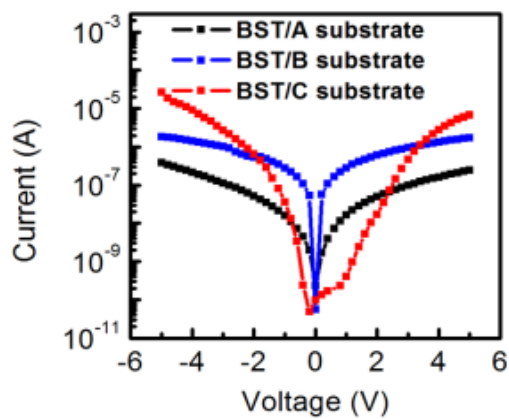

FIG. 11. Plot of leakage current versus applied voltage of BST films on different substrates.

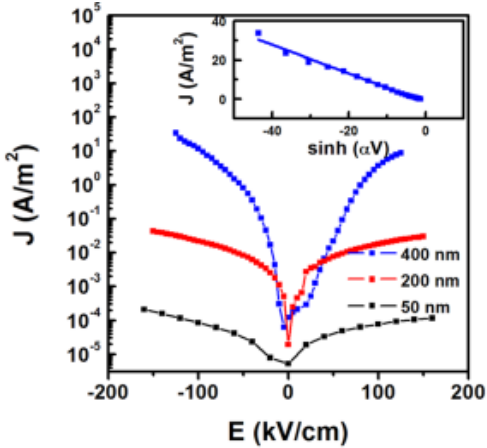

FIG. 12. The plot of J versus E of films grown on substrate C with different thicknesses. The inset shows the plot of J versus sinh(aV) for the film grown on C substrate with 400 nm thickness with linear fit.

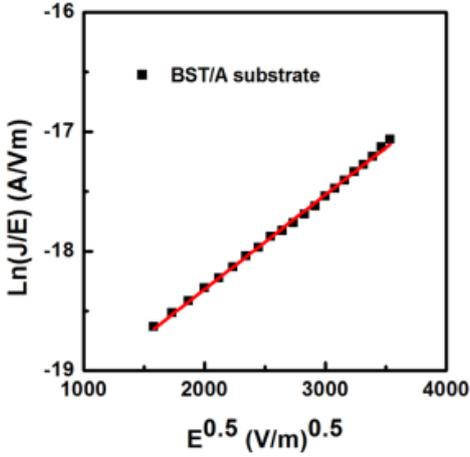

FIG. 13. The Simmons plot of ln(J/E) versus E 1/2 for the films grown on substrate A.

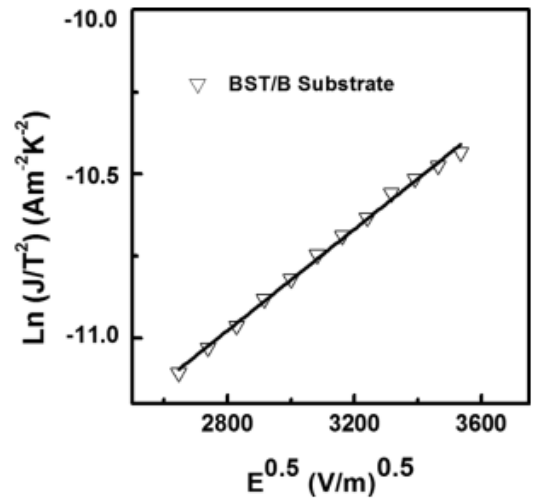

FIG. 14. The Schottky plots of ln(J/T2) versus E0.5 for the film grown on substrate B.